\documentclass[conference]{IEEEtran}
\IEEEoverridecommandlockouts

\usepackage{amsmath,amssymb,amsfonts}
\usepackage{graphicx}
\usepackage{textcomp}
\usepackage{xcolor}
\usepackage{array}
\usepackage{cite}
\usepackage{makecell}
\usepackage{balance}

\usepackage[hidelinks]{hyperref}
\hypersetup{
    pdfborder={0 0 0}
}

\usepackage[linesnumbered,ruled,vlined]{algorithm2e}
\SetKwInput{KwInput}{Input}  
\SetKwInput{KwOutput}{Output}
\SetKw{And}{and}
\SetKw{Or}{or}
\SetKw{Break}{break}
\SetKw{KwTo}{to}
\SetKw{KwDownTo}{downto}

\usepackage{tabularx}
\makeatletter
\newcommand{\multiline}[1]{%
  \begin{tabularx}{\dimexpr\linewidth-\ALG@thistlm}[t]{@{}X@{}}
    #1
  \end{tabularx}
}
\makeatother

\ifCLASSOPTIONcompsoc
\usepackage[caption=false,font=normalsize,labelfont=sf,textfont=sf]{subfig}
\else
\usepackage[caption=false,font=footnotesize]{subfig}
\fi
\begin{document}

\title{Intent-Based Orchestration in Open RAN:\\ An ns-3 Simulation Framework
\thanks{This research was funded in part by the Agence Nationale de la Recherche (ANR) under the ANR-24-IAS1-0002-02 COMSEMA project.}
% \thanks{The developed framework will be released as open-source upon publication.}
}

\author{Pouya Agheli and Grégoire Lefebvre\\Orange Research, Grenoble, France}

\maketitle

\begin{abstract}
This paper presents an extensible ns-3-based simulation framework for evaluating intent-based, semantics-aware control in Open RAN architectures. The framework integrates external Radio Access Network (RAN) Intelligent Controller (RIC) components and supports fine-grained control via internal distributed applications (dApps), enabling intent-based RAN orchestration across different timescales while maintaining standardized network behavior. As an illustrative use case, we implement an intent-based dApp for radio resource management (RRM) under realistic observability constraints. The scheduling problem is formulated using realistic key performance measurements (KPMs) available to dApps, together with a newly introduced Intent Satisfaction Score (ISS), which quantifies the delivery of intent-relevant information by combining distortion- and perception-oriented measures. Simulation results show that intent-based RRM can improve ISS while significantly reducing radio resource usage and computational overhead, at the cost of a moderate reduction in packet delivery ratio and throughput.
% This paper presents a novel, high-fidelity simulation framework for intent-based radio resource management  within the Open RAN architecture, built upon the ns-3 framework. It seamlessly integrates RAN intelligent controller components to enable the realistic modeling of semantic and intent-aware control mechanisms across the protocol stack. The proposed environment provides a practical platform incorporating AI-driven resource allocation and semantic communication, thereby advancing the development of next-generation wireless networks that are scalable, flexible, and intelligent.
% Our approach involves optimizing the scheduling process by leveraging key performance indicators, including packet delivery ratio, throughput, latency, and a novel metric termed the Intent Satisfaction Score. This metric is designed to quantitatively assess the fulfillment of predefined network intents. This optimization is conducted within the constraints imposed by available resource limitations and shows promising results.
\end{abstract}

\begin{IEEEkeywords}
Open RAN, Intent-Based RRM, Simulation
% Open RAN, RIC, MAC scheduler
\end{IEEEkeywords}

\section{Introduction}
The rapid growth of cyber-physical systems demands real-time, reliable, and resource-efficient information exchange. To scale while conserving limited communication resources, modern networks are shifting from intent-agnostic, maximalist designs toward intent-based, minimalist approaches that communicate \emph{only} information relevant to predefined intents. Under this philosophy, intent-based, semantics-aware communication has emerged as a paradigm for improving communication effectiveness and efficiency by judiciously using communication and computation resources. 

This paradigm is commonly motivated by Weaver's extension of Shannon's theory \cite{ShannonWeaver49}, which distinguishes technical, semantic, and effectiveness levels of communication. While conceptually useful, these distinctions blur in practice, where semantic representation and effectiveness-driven adaptation become tightly coupled. Consequently, the semantic and effectiveness layers can be viewed as a unified service management and orchestration layer operating alongside the traditional protocol stack. This layer coordinates application-level functions---such as message compression, feature extraction, and embedding generation---while also directing resource allocation in the Radio Access Network (RAN), enabling a full-stack redesign tailored to intent-based, semantics-aware communication. More broadly, this framework aligns with the principles of Intent-Based Networking (IBN) automation, which encompasses intent-related functions such as intent profiling, activation, and assurance \cite{leivadeas2022survey}. While conventional IBN primarily operates at higher layers and longer control-loop timescales, semantics-aware communication extends intent realization deeper into the communication process by jointly adapting application-level operations and RAN behavior.

Open RAN architecture provides a natural foundation for this vision by introducing openness, disaggregation, and programmability into the RAN. By separating hardware and software and supporting intelligent control across multiple timescales, it facilitates the integration of intent-based, semantics-aware functionality throughout the protocol stack. O-RAN Alliance specifications \cite{ORAN-arch} position the RAN Intelligence Controller (RIC) as the central platform for non-real-time (non-RT) and near-real-time (near-RT) control through modular RAN applications (rApps) and extended applications (xApps). However, the control-loop latency of the standard RIC architecture limits real-time decision-making and responsiveness to rapidly changing intents. Recent extensions \cite{d2022dapps, o2024dapps, lacava2025dapps} introduce distributed applications (dApps) that move intelligence closer to lower Open RAN nodes. This enables shorter control loops, reduced signaling overhead (up to $3.57$ times \cite{d2022dapps}), and provides granular access to local network-state information, such as per-packet information and raw I/Q samples. This distributed intelligence supports edge-optimized artificial intelligence (AI) and flexible, software-based RAN adaptation, enabling real-time, intent-based decision-making in response to network dynamics.

The ns-3 simulator provides a near-realistic simulation environment for evaluating Open RAN systems compliant with 3rd Generation Partnership Project (3GPP) standards \cite{Riley2010}, widely used for Long-Term Evolution (LTE) and 5G New Radio (NR) research. Nevertheless, ns-3 is primarily developed in C++, with RAN operations implemented as native components, which makes it difficult to integrate Open RAN elements---such as RICs and intelligent control applications. Supporting flexible and programmable Open RAN control requires either interfacing with external control modules or redesigning internal components, the latter often limiting extensibility in practice. This gap underscores the need for modular, realistic Open RAN simulation frameworks that support the incorporation of intent-based, semantics-aware orchestration.

\subsection{Related Works}
Several studies in the ns-3 and Open RAN literature have explored integrating machine learning (ML)- or AI-based external control components into network simulators, often using frameworks such as ns-O-RAN \cite{lacava2023ns}. The primary objective of these efforts is to enhance flexibility in RAN operations, especially in Medium Access Control (MAC) scheduling, through data-driven optimization. Representative examples include \cite{garey2023ran} and \cite{kim2025evolving}, which integrate external ML- and AI-empowered radio resource management (RRM) controllers with ns-3 using the ns3-gym framework \cite{ns3gym}. However, they do not support intent-based orchestration, primarily focus on proposing data-driven solutions for isolated RAN operations, and lack dApp integration, thereby limiting their applicability to near-RT or higher-layer control.

Further research explores the convergence of Open RAN and semantic communications through intelligent RRM \cite{li2023open, puligheddu2023sem2, zeydan2025semantic, elkael2025allstar}, addressing topics such as sub-millisecond scheduling, joint radio and compute slicing, RIC-enabled intelligence, and AI-driven satellite--terrestrial coordination. Despite reported gains, the broader impact remains constrained by AI-induced reasoning errors, high energy consumption, and the lack of a realistic, modular, and reproducible Open RAN simulation framework that supports intent-based, real-time control.

\subsection{Contributions}
We develop an open-source\footnote{Code available at \url{https://github.com/Orange-OpenSource/ns3-iboran}.}, reproducible simulation framework extending ns-3 for realistic evaluation and benchmarking of intent-based orchestration in Open RAN, which is released as \textbf{ib-ORAN}. The framework integrates two sets of Python-based control modules---local dApps and external RIC components---with the ns-3 network environment via a high-performance shared-memory interface, enabling coordinated control across multiple timescales while adhering to Open RAN architectural principles and 3GPP standards. Leveraging the ns3-ai module \cite{ns3ai}, the proposed design enables an efficient interaction between the C++ simulation core and Python-based control logic, achieving $50$--$100$ times lower latency \cite{ns3ai} than socket-based alternatives such as ns3-gym used in \cite{garey2023ran} and \cite{kim2025evolving}. This capability supports the realistic implementation of fine-grained, real-time RAN control mechanisms.

Using this framework, we propose an update to the Open RAN logical architecture to support intent-based orchestration, detail the resulting system structure and data flows, and implement a dApp for intent-based RRM as an initial use case for validating simulator operation.  We define an Intent Satisfaction Score to capture intent-dependent semantic relevance under practical observability constraints and formulate the corresponding RRM problem based on limited low-level measurements available to local dApps. Simulation results compare representative scheduling strategies and identify which realistic metrics best reflect intent-relevant performance.

% The proposed framework provides a practical and extensible platform for industrial research on intent-based orchestration in Open RAN and is released as an open-source ns-3 extension.

\section{Open RAN Architecture with Intent Support}
The O-RAN Alliance classifies Open RAN nodes as central, distributed, and radio units (O-CU, O-DU, and O-RU) \cite{ORAN-arch}. Although the O-RAN architecture is inherently modular and software-driven, supporting intent-based, semantics-aware control requires additional functional and interfacing elements. Addressing it involves gradually enhancing mechanisms within the service management and orchestration (SMO) framework, non-RT and near-RT RICs, and lower-layer dApp controllers. 

\begin{figure}[t!]
    \centering
    \includegraphics[width=1\linewidth]{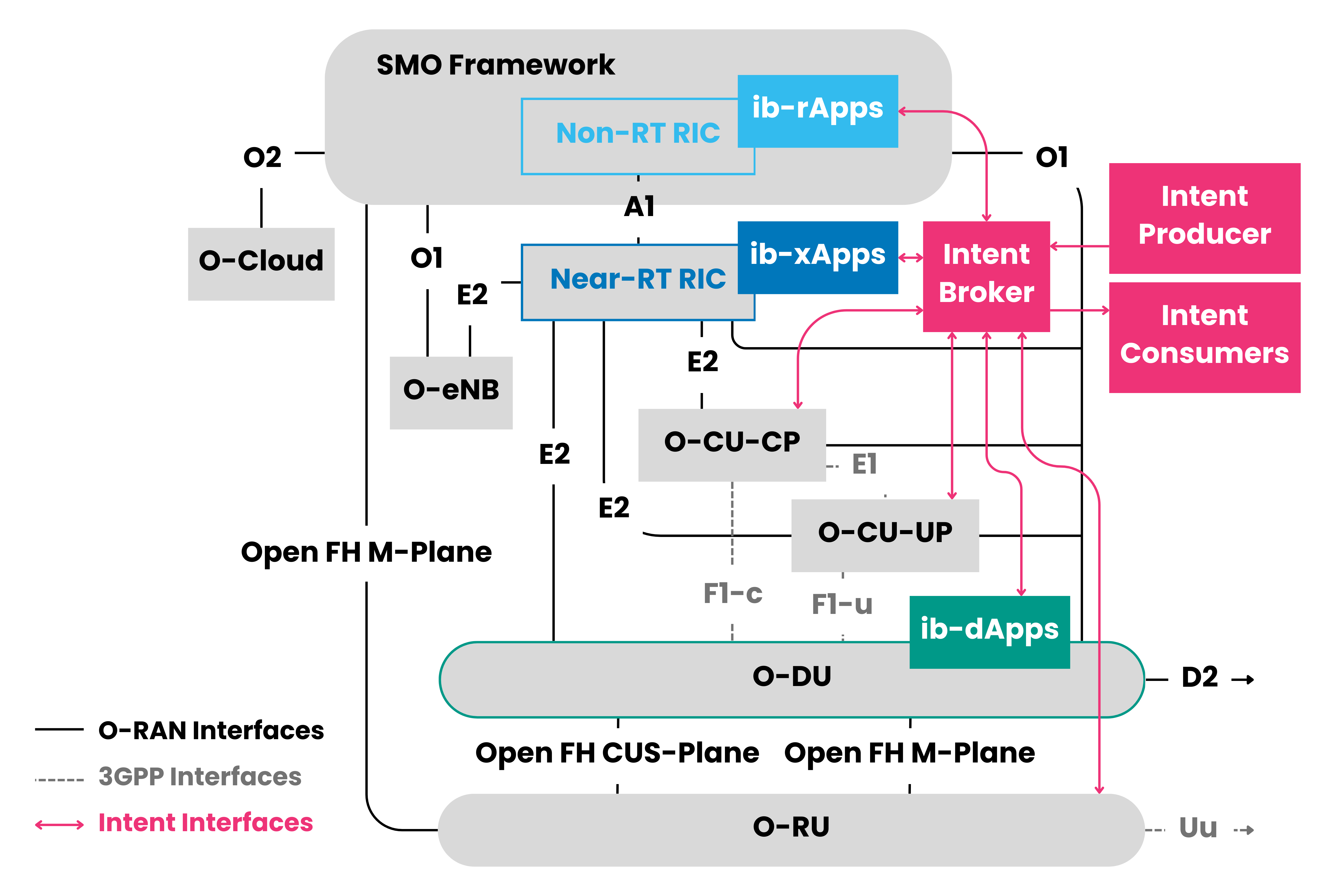}
    \caption{Open RAN logical architecture with intent support.}
    \label{fig:oran-arch}
\end{figure}

Two complementary enhancements are needed. First, new architectural elements must be defined to issue, manage, and distribute timely intents and requirements across Open RAN nodes. Second, native control components must be extended to support the execution of intent-based, semantics-aware control policies by hosting \emph{intent-based rApps (ib-rApps)}, \emph{xApps (ib-xApps)}, and \emph{dApps (ib-dApps)}. These enhancements should avoid introducing bulky components, excessive computational overhead, or disruptions to architectural modularity. Existing studies support elements of this direction (see \cite{puligheddu2023sem2} and \cite{li2023open}), though the approach proposed here distinguishes itself by integrating intent-based orchestration across multiple control layers within a unified, deployable architecture. Fig.~\ref{fig:oran-arch} illustrates the resulting architecture, which augments native Open RAN and 3GPP-compliant elements with components that explicitly support intent-based orchestration.

The \emph{Intent Broker} plays a central role in the proposed architecture. Serving as a coordination hub, it connects the Open RAN system to external entities such as the \emph{Intent Producer}, which supplies high-level intents and semantic metadata for service data units, and \emph{Intent Consumers}, which indirectly exploit this information to support coordinated operations. The Intent Broker aggregates incoming intent information, distributes global intents towards architectural layers, translates them into actionable instructions for ib-rApps and ib-xApps, and ensures consistent interpretation across control modules. It also dispatches intents directly to ib-dApps, bypassing purely syntactic processing and inter-node control signaling, thereby enabling timely and fine-grained local decision-making.

These interactions rely on dedicated \emph{Intent Interfaces} that connect intent-aware components with native elements and expose controlled access to external policy makers or consumers. The interfaces operate bidirectionally, allowing control components to provide feedback to both the Intent Broker and the Intent Producer, supporting long-term intent refinement and policy adaptation. Finally, the proposed architecture motivated the definition of intent-dependent key performance measurements (KPMs)---and consequently, key performance indicators (KPIs)---evaluated across multiple control-loop timescales to ensure that intent requirements are effectively met.

\section{Open RAN Simulation Framework}\label{sec:sim_fw}
To develop an ns-3-based simulation framework for intent-supporting Open RAN, we employ shared-memory interfaces to connect the C++ network simulation environment to two sets of Python-based control modules: one hosting ib-xApps and ib-rApps, and the other hosting ib-dApps. Each set operates over a dedicated shared-memory region. 

\begin{figure}[t!]
    \centering
    \includegraphics[width=1\linewidth]{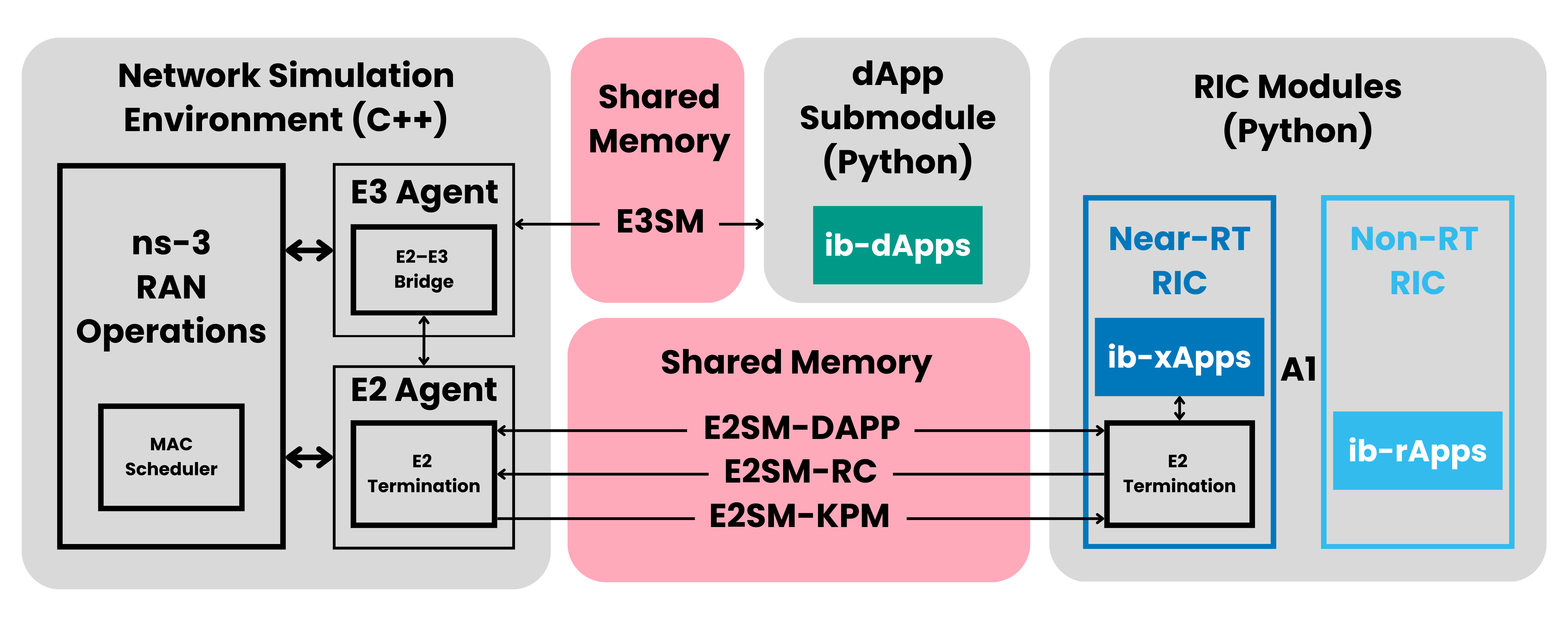}
    %\caption{Diagram showing how the network simulation environment interfaces with its control modules using a shared-memory interface.}
    \caption{Interactions between the network simulation environment and control modules via a shared-memory interface.}
    \label{fig:sim-diagram}
\end{figure}

Fig.~\ref{fig:sim-diagram} depicts the simulator's structure, highlighting its main implementation modules and supported interfaces.

\subsection{Network Simulation Environment}
We construct an evolved packet core (EPC)-enabled LTE-Advanced (LTE-A) environment using the ns-3 seminal module.\footnote{We are actively developing the simulator to support the 5G NR standard using the 5G-LENA module in the next release.} The topology comprises one or more \emph{eNodeBs} serving a group of \emph{user equipments (UEs)} over the LTE air interface. eNodeBs connect to a \emph{remote host} through the EPC core network and a high-capacity point-to-point backhaul link. End-to-end Internet Protocol (IP) connectivity is achieved by installing the Internet stack on UEs and the remote host, assigning UE IPv4 addresses via the EPC helper, and configuring routing to the EPC gateway. On the radio access side, the propagation channel encompasses both large- and small-scale effects. In addition, UEs follow a mobility model within a bounded service area, and handover is managed using a Reference Signal Received Power (RSRP)-based event-triggered mechanism.

Traffic is generated as independent uplink and downlink flows between the remote host and each active UE. In either direction, the transmitting node reads a dedicated input file or trace, segments it into fixed-size packets, and injects them into the stack based on a configurable inter-packet interval or application rate. This approach supports per-user and per-direction intent definitions, enabling the integration of application-specific requirements for every uplink and downlink flow.

According to the LTE-A specifications and the ns-3 implementation, the downlink uses cyclic-prefix orthogonal frequency-division multiplexing (CP-OFDM), while the uplink employs discrete Fourier transform (DFT)-spread OFDM, also known as single-carrier frequency-division multiple access (SC-FDMA). Service packets are carried over the Physical Downlink Shared Channel (PDSCH) and the Physical Uplink Shared Channel (PUSCH). The eNodeB MAC scheduler manages user scheduling and resource allocation per transmission time interval (TTI) lasting $1$ ms, dynamically assigning time--frequency resources to UEs in both directions while handling hybrid automatic repeat request (HARQ) operations. Control signaling, such as scheduling grants and UE feedback, is conveyed over the Physical Downlink Control Channel (PDCCH) and the Physical Uplink Control Channel (PUCCH), which are assumed to be adequately provisioned but are not explicitly optimized in this work.

By supporting the Open RAN architecture, the eNodeB MAC scheduler can be controlled either locally by ib-dApps deployed at the eNodeB dApp submodule or externally by ib-xApps (and indirectly by ib-rApps) operating within the RIC module. The MAC scheduler performs per-TTI operations based on control directives and RRM decisions issued by these decision-making components. To enable this interaction, the eNodeB E2 agent collects measurements from the MAC and Radio Link Control (RLC) layers and reports them to the near-RT RIC over the E2 interface, in accordance with the E2 Service Model for KPM (E2SM-KPM). In parallel, fine-grained state and intent-related information is delivered to the local dApp submodule through the internal E3 interface \cite{d2022dapps}, using a dedicated E3 agent compliant with the E3 Service Model (E3SM). Intent-related information retains timely intents, together with side information and criticality indicators for each UE.\footnote{Within the simulator, intent side information and criticality indicators are carried as application-layer metadata (packet tags) and exposed to the eNodeB dApp via the internal E3 interface. This approach maintains 3GPP-compliant radio procedures and avoids adding new standardized LTE signaling elements.} Decisions and policies from ib-dApps and ib-xApps are enforced within the eNodeB by updating the MAC scheduler's configuration and operational parameters.

% The key ns-3 configuration parameters initialized to model the proposed network simulation environment are summarized in Table~\ref{tab:confparams} in Section~\ref{sec:results}.

\subsection{Control Modules}
\subsubsection{dApp submodule}\label{subsec:dapp}
The local dApp submodule operates at the shortest control-loop timescale and requires fine-grained network-state information. To supply the necessary inputs for ib-dApps, the E3 agent serves as an internal application programming interface (API) endpoint that connects RAN operations, external RIC components, and the dApp submodule, providing two types of information. 

The first class consists of timely measurements from the MAC and RLC layers, collected by the E3 agent every $\Delta_{\rm dApp}$ TTIs, with $\Delta_{\rm dApp} \in [1, 10)$. To support structured information exchange, E3SM defines two complementary services. The REPORT service delivers aggregated local measurements to the dApp submodule as E3 INDICATION messages at each $\Delta_{\rm dApp}$ TTIs. Conversely, the CONTROL service enables the dApp submodule to return control actions to the E3 agent via E3 CONTROL messages, which are subsequently enforced at the MAC scheduler (see \cite{lacava2025dapps} for further details).

The second class of inputs comprises higher-level control policies and configuration directives issued by ib-xApps at a coarser timescale of $10$ TTIs.\footnote{We consider ib-xApps to operate at the minimum supported near-RT control period of $10$ ms, in line with Open RAN specifications and the discrete-event execution model of ns-3.} These directives are conveyed to the E3 agent over the E2 interface using a custom, non-standard E2 Service Model for dApp (E2SM-DAPP) \cite{lacava2025dapps}, transported via RIC CONTROL messages. An internal E2--E3 bridge enables the E3 agent to interpret these directives and forwards them to the ib-dApps. Conversely, ib-dApps generate enrichment information---such as scheduling context or allocation summaries---which is shared with ib-xApps via the same E2SM-DAPP, translated by the E2--E3 bridge into E2SM-DAPP indication payloads, and then delivered to ib-xApps as RIC REPORT messages.

To reduce signaling overhead and control-channel load, the exchange of report and control messages between the eNodeB dApp submodule and the near-RT RIC module occurs at a coarse periodicity of $10$ TTIs, since the external control loop is the primary bottleneck. Accordingly, we can define $\Delta_{\rm dApp}$ as an integer divisor of the near-RT control period. This design choice aligns fine-grained dApp updates with coarser near-RT RIC cycles, ensuring that E2- and E3-agent inputs are available at well-defined synchronization points. 

Based on the most recent local measurements, higher-level directives from ib-xApps, and time-varying intents from the Intent Broker, ib-dApps perform fine-grained decision-making every $\Delta_{\rm dApp}$ TTIs. The resulting control actions are immediately enforced by updating the MAC scheduler operation.

\subsubsection{RIC modules}
The near-RT RIC module performs coarse-grained decision-making based on two sets of input reports: higher-level measurements from the eNodeB MAC and RLC layers, and decision summaries generated by the eNodeB dApp submodule. The first set is reported over the E2 interface using the E2SM-KPM and delivered via its REPORT service as RIC REPORT messages at a periodicity of $10$ TTIs. The second set follows the E2SM-DAPP specifications (see Section~\ref{subsec:dapp}) and is conveyed in separate RIC REPORT messages that carry E2SM-DAPP-defined payloads.

Based on these report streams and timely intents from the Intent Broker, the near-RT RIC derives higher-level control policies and optimization directives every $10$ TTIs. In addition, the non-RT RIC supports near-RT decision-making by providing long-term policy guidance, optimization objectives, and, where applicable, trained AI/ML models via the A1 interface.

Coarse-grained control policies and configuration directives generated by ib-xApps are delivered to the dApp submodule over the E2 interface using the E2SM-DAPP control semantics and transported in RIC CONTROL messages. These directives are then forwarded through the internal E2--E3 bridge to ib-dApps, where they guide local control logic responsible for fine-grained MAC scheduling decisions. Consistent with the Open RAN framework, the near-RT RIC may also issue control decisions directly to RAN operations based on the E2SM-RC and its CONTROL service, delivered by RIC CONTROL messages. Such actions are enforced immediately at the MAC scheduler, bypassing longer-term policy mediation.

\begin{figure}[t!]
    \centering
    \includegraphics[width=1\linewidth]{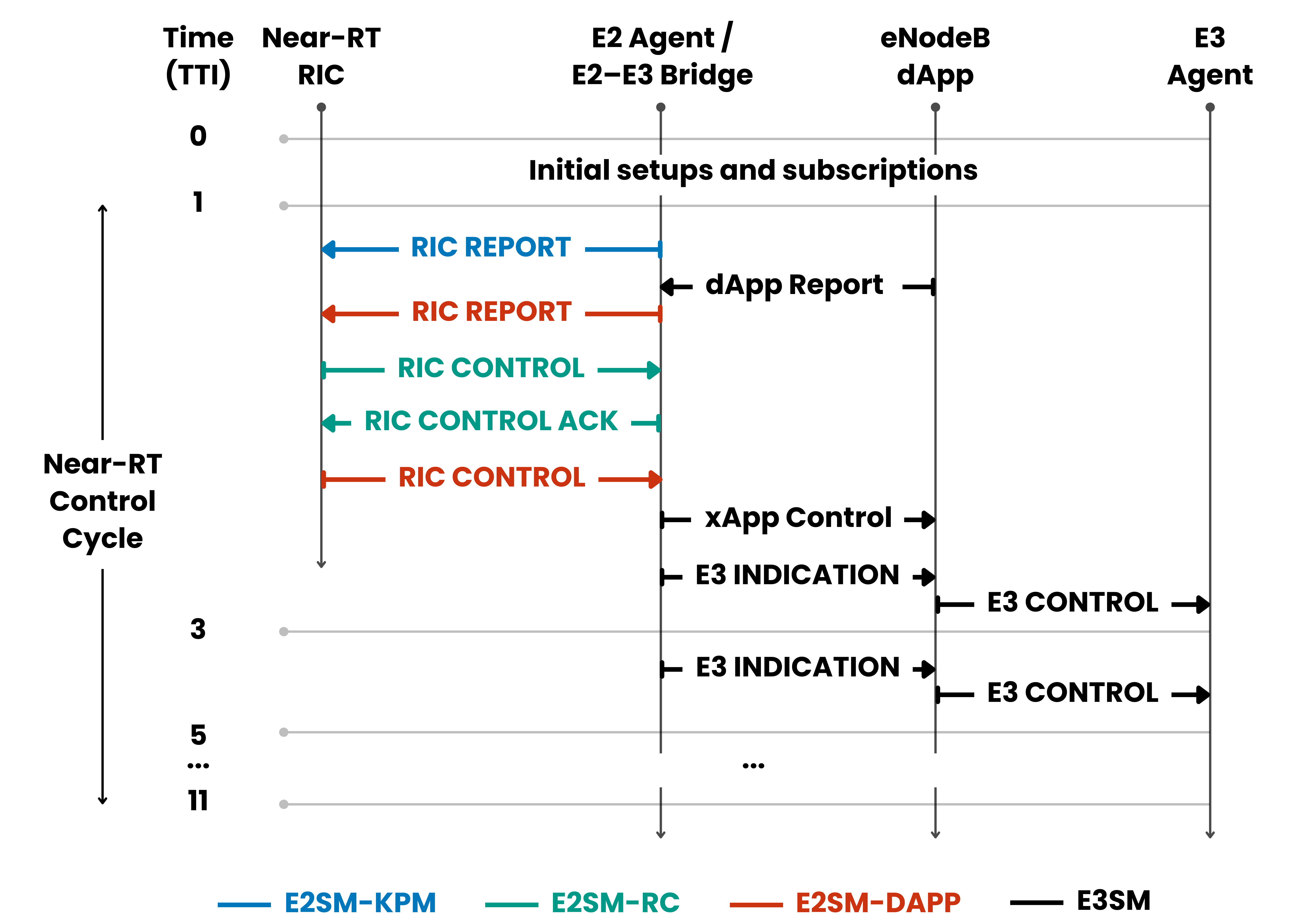}
    \caption{Time-based message exchange among the near-RT RIC, eNodeB dApp submodule, and interfaced E2 and E3 agents.}
    \label{fig:interaction}
\end{figure}

Fig.~\ref{fig:interaction} provides a time-based overview of information exchanges among simulator components during a single near-RT control cycle, detailing the associated service models and message types for each interaction. In this configuration, we set $\Delta_{\rm dApp}=2$, such that one near-RT control cycle of $10$ TTIs encompasses five consecutive local control loops. Moreover, Table~\ref{tab:message} summarizes the message payloads exchanged under each service model. Here, RNTI, CQI, QCI, and MCS denote the radio network temporary identifier, channel quality indicator, quality of service class identifier, and modulation and coding scheme, respectively.  

\noindent
\renewcommand{\arraystretch}{1.2}
\begin{table*}[t!]
    \centering
    \caption{Information exchanged over the E2 and E3 interfaces.}\label{tab:message}
    \begin{tabular}{|l|l|m{11cm}|}
    \hline
         \multicolumn{1}{|c}{\footnotesize \textbf{\!Service Model\!}} &
         \multicolumn{1}{|c}{\footnotesize \textbf{\!Message Type\!}} &
         \multicolumn{1}{|c|}{\footnotesize \textbf{Message Payload}}\\
    \hline
    \hline
         \footnotesize \!E2SM-KPM &
         \footnotesize \!RIC REPORT &
         \footnotesize Aggregated and per-UE KPMs from the MAC and RLC layers (e.g., number of active UEs, resource utilization, throughput, latency, buffer level, CQI-related indicators, etc.).\\
    \hline
         \footnotesize \!E2SM-RC &
         \footnotesize \!RIC CONTROL &
         \footnotesize High-level radio control directives and policy configurations (e.g., scheduling constraints, prioritization rules, admission control decisions, link adaptation policies, etc.).\\
    \hline
         \footnotesize \!E2SM-DAPP &
         \footnotesize \!RIC REPORT  &
         \footnotesize Total number of served UEs, allocated resources, system load handled, and per-UE information (e.g., average throughput, packet delay, total bytes served, etc.).\\
    \hline
         \footnotesize \!E2SM-DAPP &
         \footnotesize \!RIC CONTROL  &
         \footnotesize Resource budget, scheduling approaches, skipping specific frames or UEs, etc.\\
    \hline
         \footnotesize \!E3SM &
         \footnotesize \!E3 INDICATION  &
         \footnotesize Number of active UEs, available resources, current subframe number, link direction, and per-UE information (e.g., RNTI, CQI, QCI, buffer level, criticality, etc.).\\
    \hline
         \footnotesize \!E3SM &
         \footnotesize \!E3 CONTROL  &
         \footnotesize RNTI of selected UEs, per-UE resource allocation map, MCS, etc.\\
    \hline
    \end{tabular}
\end{table*}

\section{Implementation of Intent-Based RRM}
In this section, we discuss the implementation of intent-based RRM, encompassing user selection and radio resource allocation. At intervals of $\Delta_{\rm dApp}$ TTIs, the ib-dApp selects one or more UEs using a scheduling mechanism and allocates their required radio resources in accordance with high-level directives and constraints issued by ib-xApps.

The smallest allocable unit of radio resources is the Physical Resource Block (PRB). According to 3GPP LTE-A specifications, uplink transmissions require contiguous PRB allocations per carrier component due to DFT-spread OFDM, whereas contiguity is not mandatory on the downlink. In the developed simulation framework, downlink allocation follows Type-0 resource allocation, in which the system bandwidth is partitioned into Resource Block Groups (RBGs), each comprising a fixed number of PRBs determined by the total system bandwidth.

\subsection{Key Performance Indicators}\label{subsec:kpi}
To evaluate the performance of the network in handling traffic across different input data formats, the simulator supports standard KPIs, including packet delivery ratio (PDR), throughput, latency, and jitter. Additionally, it allows the definition of context- or semantics-aware metrics tailored to specific data formats and applications. Without loss of generality, in this work, both uplink and downlink traffic are generated from user-specific, labeled image files.

To quantify the extent to which network transmissions satisfy the underlying intent, we introduce the \emph{Intent Satisfaction Score (ISS)} metric, denoted by $\mathcal{S}_{\mathrm{I}}$ and defined as\footnote{To the best of our knowledge, no globally accepted metric exists for this purpose. The proposed formulation serves as a baseline, while the simulator remains agnostic to any specific ISS definition.}
\begin{align}\label{eq:iss}
    \mathcal{S}_{\mathrm{I}} &= h_{\mathrm{I}}\left(\mathcal{F} \right),~~    \mathcal{F} = \alpha f_0 + \beta f_1 + \gamma f_2,
\end{align}
where $\mathcal{F}$ is the fidelity score composed of semantic, content, and structural fidelity components, $f_0$, $f_1$, and $f_2$, respectively, each taking values in $[0, 1]$. Also, $\alpha, \beta, \gamma\in[0, 1]$ show weighting coefficients that satisfy $\alpha+\beta+\gamma=1$. The fidelity score combines distortion- and perception-oriented measures, consistent with established principles of semantic communication \cite{blau2019rethinking}. Specifically, $f_0$ captures perception-related (semantic) fidelity, $f_2$ reflects distortion-related fidelity, and $f_1$ serves as a hybrid measure that integrates both aspects.

An intent-based activation function $h_{\mathrm{I}}:[0, 1] \rightarrow [0, 1]$ maps the fidelity score $\mathcal{F}$ to the ISS based on its relevance to the underlying intent. For intent-relevant data flows with $\mathcal{F} \geq \mathcal{F}_{\rm min}$, we set $h_{\mathrm{I}}\left(\mathcal{F}\right) = \mathcal{F}$, where $\mathcal{F}_{\rm min}=0.2$ is the minimum acceptable fidelity\footnote{This threshold can be adjusted depending on the user application or intent.}; otherwise, $h_{\mathrm{I}}\left(\mathcal{F}\right)=0$.

\subsubsection{Semantic fidelity}
To derive semantic fidelity between a transmitted image and its reconstruction, we use the Contrastive Language--Image Pretraining (CLIP) model \cite{clip}. CLIP is first applied to the entire image to obtain a global similarity score. The image is then divided into equal-sized patches (in this work, $16$ patches arranged in a $4 \times 4$ grid), and semantic fidelity is evaluated for each patch separately.\footnote{Patch-level assessment captures localized semantic degradations that may not be reflected in the global similarity score.} Therefore, we have
\begin{equation}
    f_0 = \frac{g_{\rm s}}{2}\cdot\big(f_{0, \rm global} + f_{0, \rm patch}\big),
\end{equation}
where $f_{0, \rm global}$ and $f_{0, \rm patch}$ denote the global and average patch-level semantic fidelity, respectively. A spatial coverage factor $g_{\rm s}$ is also introduced, defined as the fraction of patches whose similarity score exceeds a prescribed threshold. In this work, the similarity threshold is arbitrarily set to $80\%$.

\subsubsection{Content fidelity}
To quantify content fidelity, we partition both transmitted and reconstructed images into $16$-pixel blocks and compare their intensity variance. Blocks in the transmitted image with variance above a threshold are considered content-bearing. Content fidelity is measured as the fraction of these blocks that retain comparable variance after reconstruction. A variance threshold of $10$ is adopted, which effectively distinguishes informative regions from visually flat areas in grayscale images at the considered resolution. 

\subsubsection{Structural fidelity}
Assessing structural fidelity, we employ the single-scale Structural Similarity Index (SSIM) \cite{wang2004image}, which measures local similarity between a transmitted image and its reconstruction by accounting for luminance, contrast, and structural information at the pixel level.\footnote{More advanced perceptual fidelity metrics, such as SSIMULACRA 2 \cite{ssimulacra2}, exist, but their conceptual complexity falls outside the scope of this work.}

All metrics are computed after resizing the transmitted and reconstructed images to a common resolution of $224\times224$ pixels. The content and structural fidelity components ($f_1$ and $f_2$) are evaluated on grayscale images.

\subsubsection{Intent definition}
In this work, intents correspond to requests for specific object identities (IDs) within a labeled image dataset. Each image is annotated with the IDs of the objects it contains, which may include one or more. During every simulation episode, each UE randomly selects an image for uplink transmission while the remote host independently selects images for downlink transmission to the UEs. At the beginning of the episode, the Intent Producer randomly activates one object ID as the current intent. The fidelity score $\mathcal{F}$ is therefore deemed intent-relevant only if the transmitted image contains the requested object ID. Otherwise, the transmission does not contribute to the ISS through  $h_{\mathrm{I}}\left(\mathcal{F}\right)$, regardless of its visual fidelity.

\subsection{Problem Formulation}\label{subsec:probform}
A natural approach to designing the ib-dApp for RRM is to formulate an optimization problem that maximizes a weighted combination of standard KPIs and the ISS through long-term, end-to-end objectives. However, such a formulation rests on strong assumptions about the stability of intent and the feasibility of defining a system-level objective for a low-level control loop operating under real-time, time-varying network dynamics. In practice, these assumptions are difficult to meet. Furthermore, in an ns-3 implementation compliant with 3GPP standards, the eNodeB dApp submodule primarily observes fine-grained network-state measurements from the MAC and RLC layers at the eNodeB and the UEs. Since intent satisfaction can only be evaluated after complete image transmission and reconstruction, the ib-dApp cannot access the fidelity score---and hence the ISS---in real time.

As a result, RRM decisions rely on information conveyed through E3 INDICATION and xApp Control messages (see Section~\ref{sec:sim_fw}), together with intent-relevance indicators such as side information and context-aware criticality. Under these constraints, the RRM objective combines intent relevance with available low-level, real-time KPMs, including CQI, QCI, and buffer level. We therefore adopt a grant-based, buffer-aware resource allocation strategy in which each selected UE is assigned sufficient downlink or uplink radio resources to drain its buffer during its scheduled TTIs. The required number of PRBs (or RBGs) is determined from the UE’s CQI, MCS, and buffer level, looking up predefined mapping tables specified in the 3GPP LTE-A standards \cite{3gpp_ts_136213}.

Consequently, the RRM problem reduces to a user-selection problem driven by the estimated resource demands, low-level KPMs, and intent-relevance indicators observable at the dApp submodule. We formulate the resulting scheduling task as a classical knapsack problem \cite{knapsack}:
\begin{align}\label{eq:opt}
    \mathcal{P} :~ &\underset{\{x_i\in \{0, 1\}
    \}_{i=0}^n}{\operatorname{max}} ~~ \sum_{i= 1}^{n} v_i(\kappa_i)x_i \nonumber \\
    & {\rm s.t.} ~~ \sum_{i= 1}^{n} b_ix_i \leq B_{\rm max},
\end{align}
where $n\geq0$ shows the number of active UEs (i.e., UEs with pending data) during a scheduling interval of $\Delta_{\rm dApp}$ TTIs. Also, $b_i>0$ denotes the estimated number of PRBs (or RBGs) required by the $i$-th UE for uplink (or downlink) transmission, and $B_{\rm max}$ represents the total radio resources available for \emph{new} transmissions\footnote{A portion of resources is dynamically reserved for HARQ retransmissions.} in the current scheduling interval.

Moreover, $v_i$, $\forall i$, is the scheduling utility associated with selecting the $i$-th UE. This utility is determined from a chosen KPM, denoted by $\kappa_i$, which depends solely on information available to the ib-dApp, such as intent-relevance indicators and low-level measurements (e.g., CQI, QCI, and buffer level). In Section~\ref{sec:results}, we identify which uplink and downlink KPMs best align with ISS-related performance under the given observability constraints.\footnote{Although no single KPM fully captures intent satisfaction, this analysis provides practical guidance on proxy metrics that best correlate with ISS under realistic deployment conditions.} 

\subsection{Scheduling Algorithm}\label{subsec:algos}

Regardless of the specific KPM used to model the scheduling utility, we tailor a greedy density-based UE selection algorithm to solve problem $\mathcal{P}$. Each UE $i$ is assigned a utility density $\rho_i=v_i(\kappa_i) / b_i$, defined as the ratio between its utility and estimated radio resource requirement. Active UEs are sorted in descending order of $\rho_i$ and selected iteratively until the available resources are depleted. This algorithm exhibits time and space complexities of $\mathcal{O}(n\log n)$ and $\mathcal{O}(n)$, respectively, enabling low-latency decision-making with reduced computational overhead compared to conventional alternatives such as dynamic programming or exhaustive combinatorial search. These properties make the greedy density-based strategy suitable for real-time RRM in ib-dApps. Nonetheless, the method is generally suboptimal for the formulated knapsack problem and achieves optimality only in the special case where all UEs require the same amount of resources.

\section{Simulation Results}\label{sec:results}
In this section, we validate the performance of the developed ns-3-based Open RAN simulator, compliant with 3GPP LTE-A specifications, and investigate the functionality of the designed ib-dApp for RRM. We compare multiple UE selection strategies using different KPMs and intent-relevance indicators available within the eNodeB dApp submodule. We then identify which KPM and corresponding scheduling utility definition (see Section~\ref{subsec:probform}) best align with the ISS-based evaluation for both uplink and downlink. In this evaluation, we set $\alpha=0.4$ and $\beta=\gamma=0.3$ to specify $\mathcal{S}_{\mathrm{I}}$ in \eqref{eq:iss}. The results are averaged over $\Delta_{\rm dApp} \in \{1, 4, 8\}$.

\noindent
\renewcommand{\arraystretch}{1.2}
\begin{table}[t!]
    \centering
    \caption{ns-3 configuration parameters.}
    \label{tab:confparams}
    \begin{tabular}{|m{2.08cm}|c||m{2.23cm}|c|}
    \hline
       \footnotesize \textbf{\!\!Parameter\!\!} & \footnotesize \textbf{\!\!Value\!\!} & \footnotesize \textbf{\!\!Parameter\!\!} & \footnotesize \textbf{\!\!Value\!\!} \\
    \hline
       \footnotesize \!\!Simulation area\!\! & \footnotesize \!\!\!$500\!\times\!500$ $\text{m}^2$\!\!\! &        \footnotesize \!\!Simulation episode\!\!  & \footnotesize \!\!$10$ s\!\! \\
    \hline
       \footnotesize \!\!Number of UEs\!\!  & \footnotesize \!\!$10$\!\! & 
       \footnotesize \!\!Number of eNodeBs\!\! & \footnotesize \!\!$1$\!\! \\
    \hline
       \footnotesize \!\!Uplink packet size\!\!  & \footnotesize \!\!\!$1400$ bytes\!\!\! & 
       \footnotesize \!\!Downlink packet size\!\!  & \footnotesize \!\!$1400$ bytes\!\! \\
    \hline
       \footnotesize \!\!Uplink source rate\!\! & \footnotesize \!\!$100$ kbps\!\! &
       \footnotesize \!\!Downlink source rate\!\! & \footnotesize \!\!$200$ kbps\!\! \\
    \hline
        \footnotesize \!\!Uplink L4 protocol\!\! & \footnotesize \!\!UDP\!\! &
       \footnotesize \!\!Downlink L4 protocol\!\! & \footnotesize \!\!UDP\!\! \\
    \hline
       \footnotesize \!\!Uplink EARFCN*\!\!  & \footnotesize \!\!$20750$\!\! & 
       \footnotesize \!\!Downlink EARFCN\!\! & \footnotesize \!\!$2750$\!\! \\
    \hline
       \footnotesize \!\!Uplink PRBs\!\!  & \footnotesize \!\!$100$\!\! & 
       \footnotesize \!\!Downlink RBGs\!\! & \footnotesize \!\!$25$\!\! \\
    \hline
       \footnotesize \!\!Downlink RBG size\!\!  & \footnotesize \!\!$4\,\text{PRB}$\!\! & 
       \footnotesize \!\!System bandwidth\!\! & \footnotesize \!\!\!$20$ MHz\!\!\! \\
    \hline
       \footnotesize \!\!Per-PRB bandwidth\!\!  & \footnotesize \!\!$180$ kHz\!\! & 
       \footnotesize \!\!Per-RBG bandwidth\!\! & \footnotesize \!\!\!$720$ kHz\!\!\! \\
    \hline
       \footnotesize \!\!LTE-A TX mode\!\!  & \footnotesize \!\!TM$2$ \cite{3gpp_ts_136213}\!\! & 
       \footnotesize \!\!P0 nominal PUSCH\!\! & \footnotesize \!\!\!$-96$ dBm\!\!\! \\
    \hline
       \footnotesize \!\!UE TX power\!\!  & \footnotesize \!\!$[-40, 23]$ dBm\!\! & 
       \footnotesize \!\!eNodeB TX power\!\! & \footnotesize \!\!\!$46$ dBm\!\!\! \\
    \hline
        \multicolumn{4}{l}{\scriptsize \!\!*EARFCN: 3GPP-defined index identifying the LTE-A carrier frequency \cite{3gpp_ts_36101}.\!\!} \\
    \end{tabular}
\end{table}
We consider a scenario with $10$ UEs initially distributed at random within a $500\times500$ $\text{m}^2$ area, connected to a single eNodeB with a tower height of $30$ m. Each UE has a height of $1.5$ m and moves at a constant speed of $3$ km/h under a two-dimensional random walk mobility model, changing direction every $10$ m. Large-scale propagation is modeled using the HybridBuildings propagation loss model \cite{ns3_hybridbuildings} in a medium-sized urban environment, while small-scale fading follows a 3GPP-compliant trace-based Extended Pedestrian A (EPA) channel model \cite{3gpp_tr_36814}.
Unless otherwise specified, the key ns-3 configuration parameters and their values are listed in Table~\ref{tab:confparams}; all other parameters follow the default ns-3 settings.

Applying the proposed greedy density-based UE selection algorithm in Section~\ref{subsec:algos}, the utility density is defined as a function of measurements available to the dApp submodule, including CQI, buffer level, and context-aware criticality\footnote{The criticality of a UE is determined by the importance of its pending packets. Each packet corresponds to a portion of an image compressed in the Portable Network Graphics (PNG) format and may contain chunks of varying importance. According to the PNG specification, the image header (IHDR) is assigned the highest criticality, followed by the palette (PLTE) and the first image data chunk (IDAT), with decreasing criticality for subsequent IDAT chunks. The image end (IEND) chunk has the lowest criticality.}, combined with proportional fairness computed from the UEs’ instantaneous achievable data rate and historical throughput. Each utility definition yields a distinct greedy selection strategy. Additionally, conventional round-robin scheduling is adopted as a baseline. For each strategy, an intent-based variant (denoted IB) is evaluated. At each control interval, UEs are partitioned into relevant and irrelevant groups based on whether their content object ID matches the requested intent. Relevant UEs are unconditionally admitted to the candidate pool, after which irrelevant UEs are greedily admitted in order until the total resource demand reaches the available capacity.

\begin{figure}
    \centering
    \subfloat[ISS metric]{
        \centering
        \includegraphics[width=\linewidth]{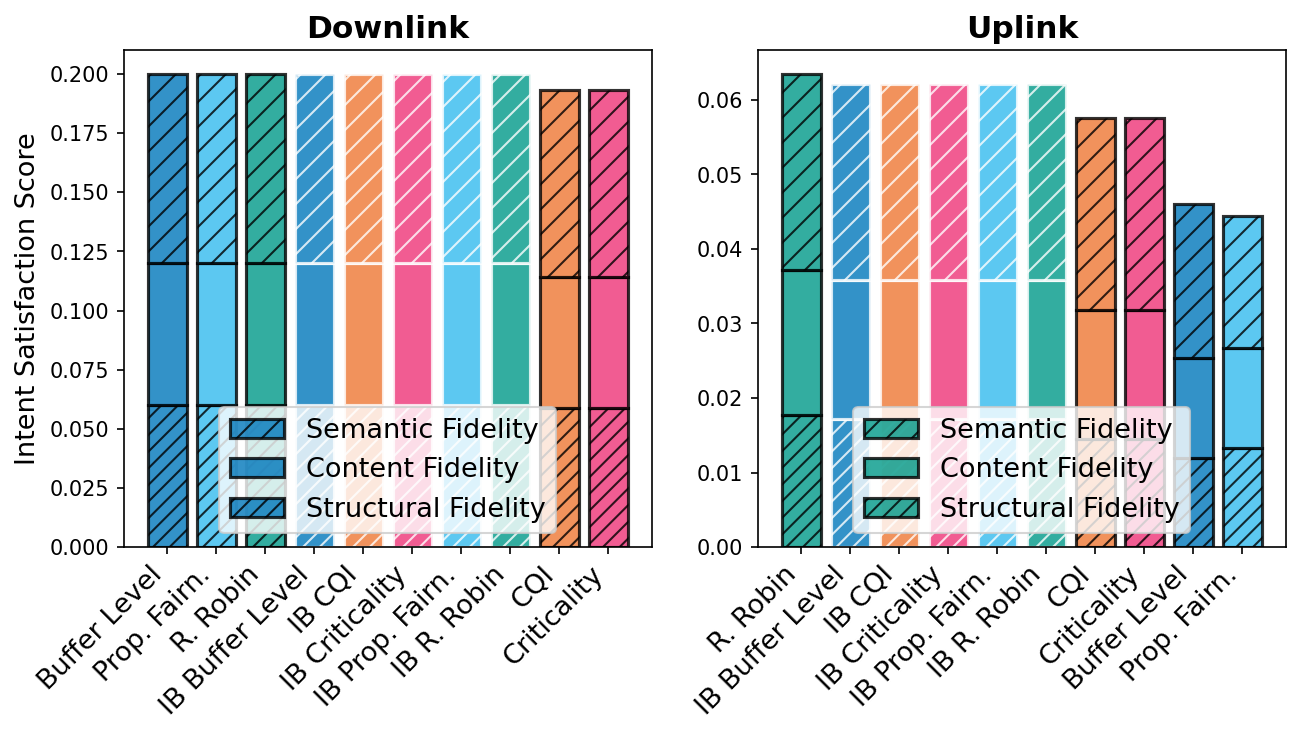}
        \label{fig:iss}
    }
    \vspace{-0.15cm}
    \hfil
    \subfloat[PDR metric]{
        \centering
        \includegraphics[width=\linewidth]{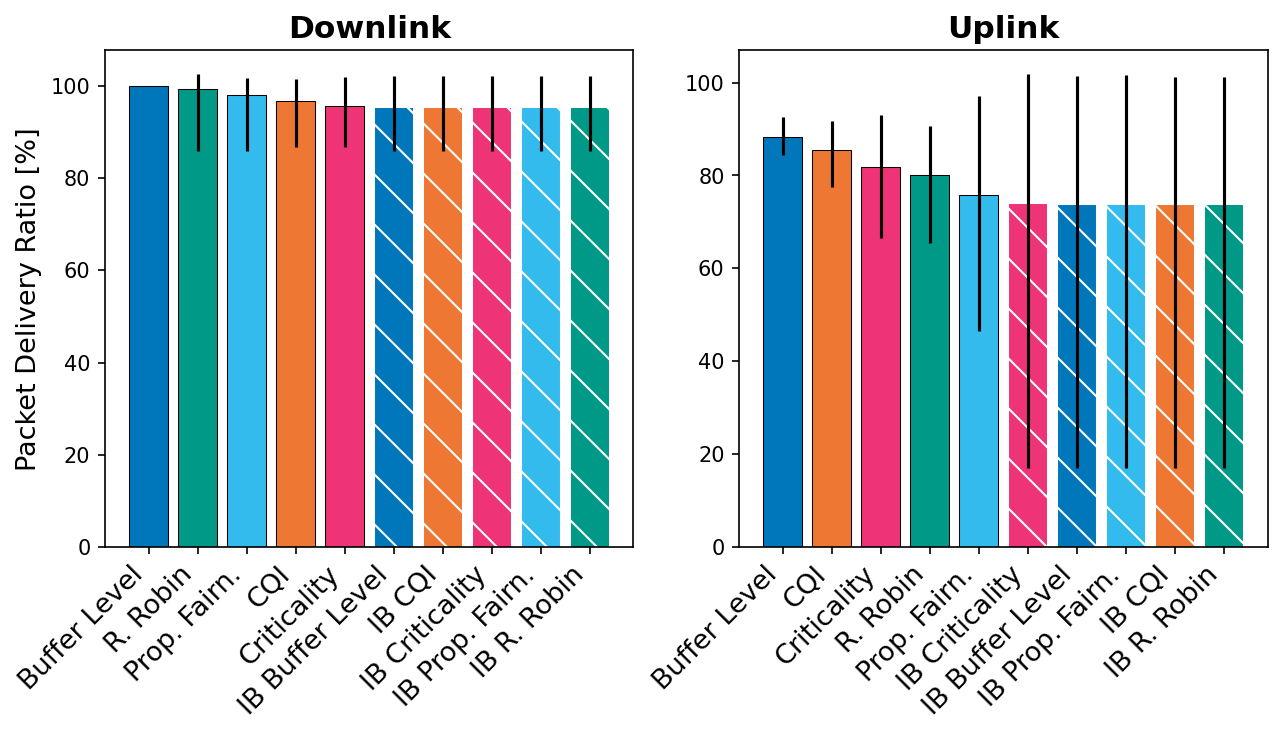}
        \label{fig:pdr}
    }
    \vspace{-0.15cm}
    \hfil
    \subfloat[Throughput metric]{
        \centering
        \includegraphics[width=\linewidth]{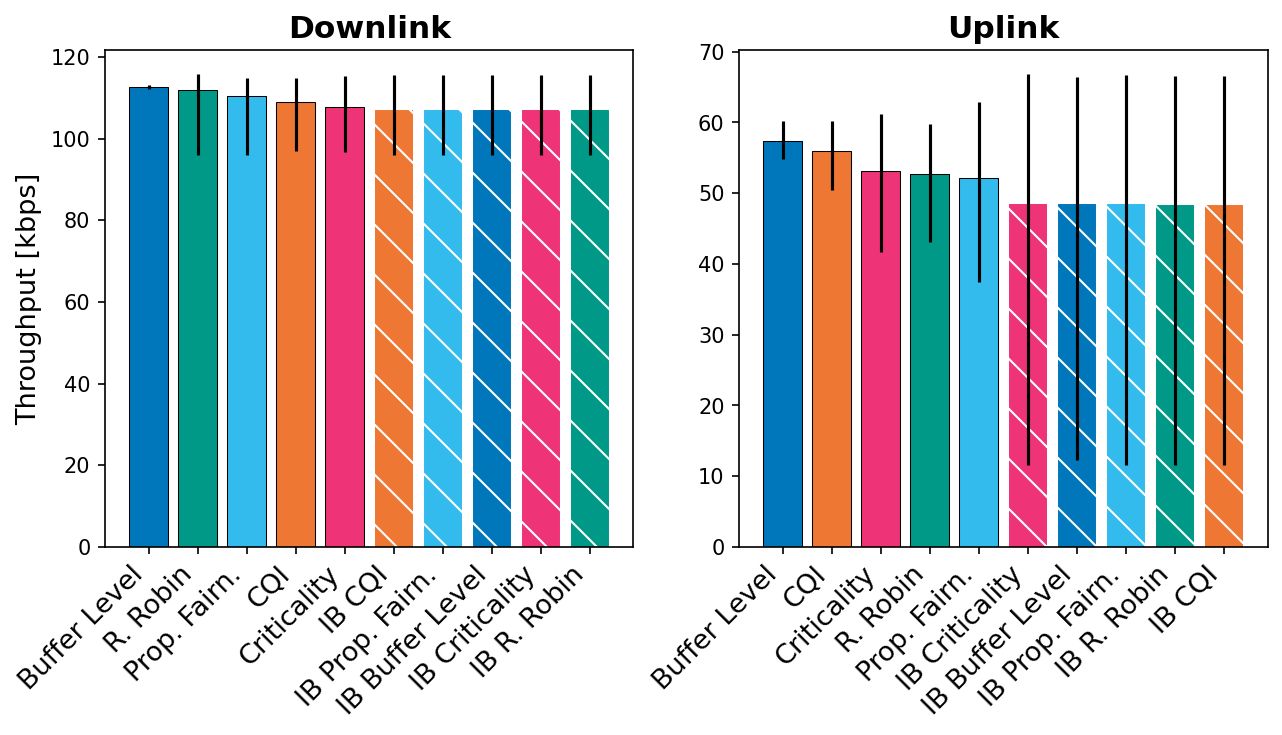}
        \label{fig:throughput}
    }
    \vspace{-0.15cm}
    \hfil
    \subfloat[Latency metric]{
        \centering
        \includegraphics[width=\linewidth]{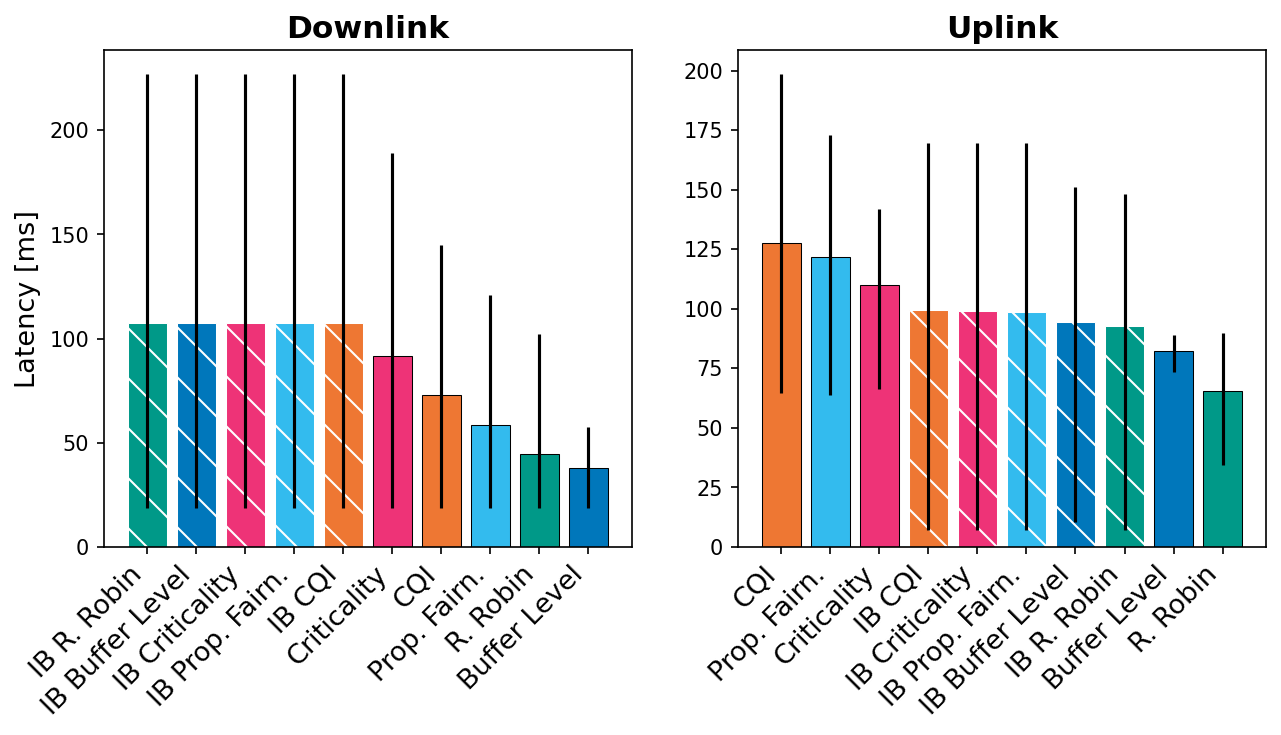}
        \label{fig:latency}
    }
    \caption{Comparison of intent-based (IB) and intent-agnostic UE selection strategies across multiple KPIs.}
    \label{fig:pm}
\end{figure}

\begin{figure}
    \centering
    \includegraphics[width=0.49\linewidth]{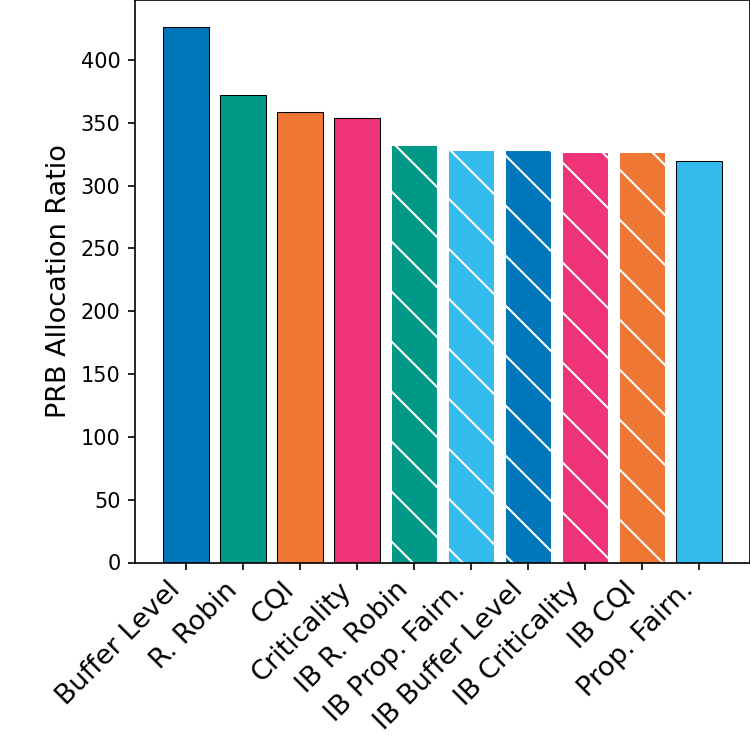}
    % \hfill
    \includegraphics[width=0.49\linewidth]{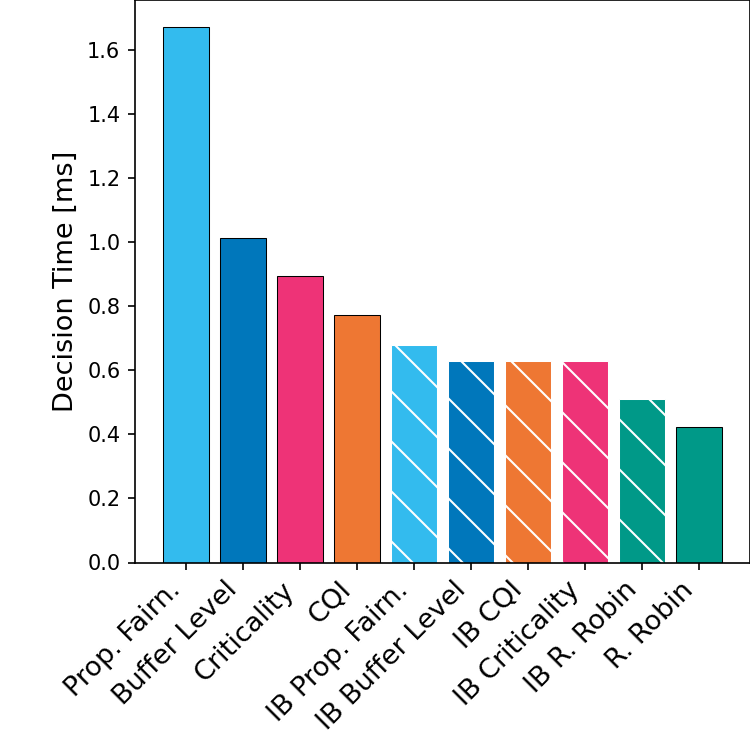}
    \vspace{-0.5cm}
    \caption{Average resource usage and decision time under different UE selection strategies across uplink and downlink.}
    \label{fig:sm}
\end{figure}

Fig.~\ref{fig:pm} compares Open RAN performance across different intent-based and intent-agnostic UE selection strategies for the formulated RRM problem (see Section~\ref{subsec:probform}). Performance is evaluated based on the ISS metric and conventional KPIs---PDR (packet delivery ratio), throughput, and latency---for both downlink and uplink transmissions (see Figs.~\ref{fig:iss}--\ref{fig:latency}). Furthermore, to highlight the associated costs, Fig.~\ref{fig:sm} reports the average percentage of allocated PRBs and the decision time for each strategy across both uplink and downlink.

For both downlink and uplink transmissions, intent-based approaches achieve higher average intent satisfaction than their intent-agnostic counterparts, albeit at the cost of reduced PDR and throughput, as UEs are prioritized based on information relevance rather than on instantaneous network conditions. According to Fig.~\ref{fig:pm}, applying intent-based UE selection increases average intent satisfaction by up to $3.3\%$ ($28.48\%$) on the downlink (uplink), while reducing PDR and throughput by an average of $2.75\%$ ($11.3\%$) and $2.8\%$ ($11.8\%$), respectively.

Additionally, intent-based approaches yield a $4.1\%$ reduction in average uplink transmission latency. The primary reason is that UEs carrying relevant information are consistently prioritized within a simulation episode, ensuring timely and reliable access to transmission and retransmission opportunities, even under unfavorable channel conditions. As a result, these UEs experience reduced waiting times between successive transmission attempts. In contrast, under intent-agnostic strategies, mobile UEs may experience intermittent resource allocation due to variations in channel quality or buffer levels across multiple intervals, leading to increased latency before successful packet delivery. Nonetheless, in the downlink, the higher transmission power at the eNodeB and the lower loss probability shift the dominant factor from channel limitations to scheduling effects. In this case, intent-based prioritization increases latency by $43.1\%$ on average, as UEs with irrelevant information may encounter long queuing delays.

Moreover, by prioritizing UEs with intent-relevant information, intent-based strategies allocate resources more selectively. Although UEs with poor channel conditions may require more resources individually, overall resource usage is reduced by avoiding allocations to less important UEs, leading to an average $8.2\%$--$30\%$ reduction in resource consumption (see Fig.~\ref{fig:sm}) while achieving higher intent satisfaction (see Fig.~\ref{fig:iss}). This efficiency is further reflected in an average $51\%$ reduction in decision time, mainly due to pre-selection and the smaller set of relevant candidates processed at each control interval.

\section{Conclusions and Perspectives}
This work presents an open-source, modular ns-3-based simulation framework for the realistic evaluation of intent-based, semantics-aware control in Open RAN architectures. By seamlessly integrating RIC components and enabling fine-grained, real-time control through dApps compliant with 3GPP LTE-A specifications, the framework supports systematic experimentation with intent-based orchestration across multiple timescales. As a use case, we implement an intent-based dApp for RRM and introduce the Intent Satisfaction Score (ISS) metric to quantify the delivery of intent-relevant information by combining distortion- and perception-oriented measures.

Simulation results demonstrate the framework’s capability to evaluate multidimensional performance across intent satisfaction, reliability, throughput, latency, and radio resource utilization under realistic observability constraints. The results show that intent-based RRM improves ISS while significantly reducing radio resource utilization and computational overhead, at the expense of relatively lower PDR and throughput and increased downlink latency. These findings expose inherent trade-offs between intent-dependent effectiveness and conventional network performance metrics, underscoring the importance of aligning scheduling design with deployment-specific objectives, target KPIs, and operational limitations.

Looking ahead, future research will focus on extending the framework to support 5G NR standards, incorporating predictive ML/AI-based control mechanisms, and exploring broader applications such as multi-tenant slicing. Further validation in real-world environments and the development of standardized intent-aware metrics will be essential steps toward the practical deployment of intent-based, semantics-aware control in Open RAN systems. More broadly, the framework contributes to the development of intelligent, autonomous, and semantics-aware wireless systems envisioned in emerging 6G architectures.

\bibliographystyle{IEEEtran}
\bibliography{references.bib}

\end{document}